\documentclass[a4paper]{VisionStyle}
\usepackage{epsfig}

\begin{document}

\title{STUDY OF X-RAY REFLECTION FEATURES IN A SAMPLE OF SY 1S 
OBSERVED BY BeppoSAX}

\author{A.\,De Rosa\inst{1,2} \and A.C.\,Fabian\inst{1}  \and L.\,Piro\inst{2} \and D. R.\,Ballantyne\inst{1}} 

\institute{
  Institute of Astronomy, Cambridge, UK.
\and Istituto di Astrofisica spaziale, Roma, Italy.
}

\maketitle 

\begin{abstract}
The ionized disc reflection model of Ross \& Fabian is applied to a 
sample of five Seyfert 1 galaxies observed by BeppoSAX. 
Good fits were obtained for all the objects in the sample.
NGC~4051 and Mkn~335 (usually classified as NLSy 1s) 
show evidence of a highly ionized accretion disc 
($\xi \sim$ 350 erg cm s$^{-1}$ and $\xi \sim$ 800 erg cm s$^{-1}$).
While NGC~3783 seems to be a ``pure Sy 1'' characterized by a highly ionized 
accretion disc ($\xi \sim$ 500 erg cm s$^{-1}$). 
All the sources are consistent with having a face-on disc. 
The accurate determination of the continuum 
emission allows us to investigate the spectral variability in the case 
of Mkn 509.   

\end{abstract}

\section{Introduction} 

Evidence for ionized accretion discs in Sy 1s  was found in recent 
observations by XMM and XMM-BeppoSAX (Mkn 509, \cite{aderosa-C2:pounds01};
Mkn 205, \cite{aderosa-C2:reeves01}; NCG 5506, \cite{aderosa-C2:matt01}). 
This conclusion was reached by considering the energy of the Fe K$\alpha$ 
line. 
The ionized disc also affects the whole continuum 
(\cite{aderosa-C2:rf93}, hereafter RF, 
\cite{aderosa-C2:ballantyne01b}). The wide energy range of BeppoSAX 
(0.1-200 keV) and its moderate spectral resolution 
($\Delta$E/E ~ 8\% at 6 keV)
allow us to measure the intrinsic continuum with good accuracy and 
to simultaneously observe both the Fe emission line and the 
Compton reflection hump. 
We present  preliminary results of fitting BeppoSAX spectra of a 
sample of Seyfert 1 galaxies with the ionized disc model of RF.
\begin{table*}
\caption{The Sample}
\begin{flushleft}
\begin{tabular}{l c c c c c c c}
\noalign{\hrule}
\noalign{\medskip}
Source & Date & Redshift (z) & \bf N$_H^{Gal}$ & cts (2-10 keV) &  
t$_{exp}$ (s) & F$^{obs}_{(2-10 keV)}$ & Ref \\
& & & (in $10^{20}$ cm$^{-2}$) & (s$^{-1}$) & (MECS) & (erg cm$^2$ s$^{-1}$) & \\
\hline
Mkn~335 & 1998-Dec-10 & 0.025 & 3.7 & 0.090 $\pm$ 0.001 & 86637 & 0.8$\times 10^{-11}$ & \cite{aderosa-C2:bianchi01} \\
Mkn~841 & 1999-Jul-30 & 0.036 & 2.2 & 0.130 $\pm$ 0.001 & 87793 & 1.3$\times 10^{-11}$ & \cite{aderosa-C2:bianchi01} \\
NGC~3783 & 1998-Jun-6 & 0.009 & 9.6 & 0.663 $\pm$ 0.001 & 153870 & 6.0$\times 10^{-11}$ & \cite{aderosa-C2:me} \\
NGC~4051 & 1998-Jun-28 & 0.002 & 1.3 & 0.203 $\pm$ 0.004 & 11616 & 1.9$\times 10^{-11}$ & This proceeding \\
Mkn~509 & 1998-May/Oct & 0.034 & 4.4 & 0.603 $\pm$ 0.004 & 87762 & 5.6$\times 10^{-11}$ & \cite{aderosa-C2:perola00} \\
Mkn~509 & 2000-Nov-3/24 & 0.034 & 4.4 & 0.278 $\pm$ 0.002 & 152380 & 2.6$\times 10^{-11}$ & This proceeding \\
\hline
\end{tabular}
\end{flushleft}
\label{aderosa-C2_journal}
\end{table*}
\section{The selected sources}

Details of of BeppoSAX observations and the sources are shown in Table 
\ref{aderosa-C2_journal}. 
The sample includes observations of sources with F(2-10 keV) greater than 
~10$^{-11}$ erg cm$^{-2}$ s$^{-1}$ and 
exposure time greater than 50 ks (with the exception of NGC~4051). This 
allowed us to obtain a good signal in the PDS band. We search all the 
sources for spectral variability. We accepted a source as 
variable if the probability P($\chi^2$) that either the 
MECS (5-10  keV)/MECS (3-5 keV)  or PDS (13-200 keV)/MECS(5-10 keV) 
hardness ratio was constant was $<$ 0.01. This allowed us to merge 
together the four short observations of Mkn~509 in 2000.
The observation of this source in 1998 had a flux higher by a factor 
of 2 from the observations performed in 2000, so it is analysed separately.
NGC~3783 also showed modest spectral variability, 
but the strength of the variations in that source is 
not relevant to the present work
(see \cite{aderosa-C2:me} for details about spectral variability). 
Therefore, we analysed the averaged spectrum extracted by the 
total observation.

\begin{table*}
\caption{Results I: The ionized disc reflection model. The disc emissivity 
law is $\epsilon(r)\propto r^\beta$, with $\beta$=-3. The inner radius is 
kept fixed: r$_{in}$=6 r$_g$. with r$_g$=Gm/c$^2$ the gravitational radius.}
\begin{flushleft}
\begin{tabular}{l c c c c c c }
\noalign{\hrule}
\noalign{\medskip}
Source & \bf $\Gamma$ & log ($\xi$) & $R$ & r$_{out}$(r$_g$) & $i$(degrees) & $\chi^2$/dof \\
& & & & & &\\
\hline
$^\bullet$Mkn~335 & 2.08$\pm^{0.06}_{0.04}$ & 2.9$\pm^{0.3}_{0.1}$ & 2.1$\pm^{0.7}_{0.5}$ & $^\star$6.9 & 29.5$\pm^{2.1}_{2.6}$ & 97.8/94  \\
Mkn~841 & 1.97$\pm^{0.03}_{0.04}$ & 1.38$\pm^{0.24}_{0.28}$ & 1.57$\pm^{0.27}_{0.22}$ & $>$40 & $<$20 & 115/109  \\
$^\bullet$NGC~3783 & 1.70$\pm0.01$ & 2.73$\pm^{0.04}_{0.05}$ & 0.49$\pm^{0.06}_{0.04}$ & $^\star$6.9 & 35.6$\pm^{2.2}_{1.1}$ & 104/101 \\
$^\bullet$NGC~4051 & 1.69$\pm^{0.04}_{0.06}$ & 2.65$\pm^{0.19}_{0.09}$ & 1.18$\pm^{0.35}_{0.26}$ & 7$\pm^{4}_{1}$ & 23.5 $\pm^{6.5}_{17.2}$ & 99.5/106  \\
Mkn~509 (1998) & 1.90$\pm$0.01 & 1.36$\pm^{0.19}_{0.14}$ & 0.83$\pm^{0.10}_{0.22}$ & 34$\pm$26 & 32$\pm$10 & 120/143 \\
Mkn~509 (2000) & 1.72$\pm${0.02} & 1.43$\pm${0.19} & 1.44$\pm^{0.20}_{0.23}$ & $^\star$10 & 11$\pm^{27}_3$ & 128/143  \\
\hline
\end{tabular}
\end{flushleft}
\small{$^\bullet$ For these sources the parameters are the 
best fit model has to take into account the complex soft X-rays spectra.
A warm absorber component is therefore added to the RF model 
(see discussion in Section \ref{aderosa-C2_discussion}).}
\newline
\small{$^\star$ Parameter fixed at that value in the fit.}
\label{aderosa-C2_iondisctable}
\end{table*}

\section{The Ionized disc model}

The ionized disc model we used to fit the spectra is described in 
RF and  \cite*{aderosa-C2:ross99}. The most important quantity in determining 
the shape of the reflected continuum is the ionization parameter 
$\xi=4\pi F_x/n_H$, 
where F$_x$ is the X-ray flux (between 0.01-100 keV) illuminating a slab of 
gas with solar abundances and constant hydrogen number density 
n$_H$=10$^{15}$ cm$^{-3}$. The incident flux is assumed be a power law 
with spectral photon index $\Gamma$ and a high energy cut-off at 100 keV. 
The computed reflected spectrum is multiplied for a factor 
$R$ (reflected fraction) 
and added to the primary continuum. The Fe K$\alpha$ emission line is also 
included in the model. Larger values of $\xi$ indicate a more ionized disc 
and this will affect the strength and width of the Fe line 
(\cite{aderosa-C2:matt93}, \cite{aderosa-C2:matt96}), and the 
absorption edges. 
The soft emission from the accretion disc is not included in the models.

\section{Spectral fitting}

Each source spectrum was fitted with the RF model described in the 
preview section (the best fit parameters are listed in Table 
\ref{aderosa-C2_iondisctable}). 
We fitted LECS (0.15-3 keV), MECS (1.5-10 keV) and PDS (13-200 keV) data
simultaneously. 
The cold absorption is kept fixed to the galactic value. Relativistic 
smearing was also taken into account by blurring the model with a kernel 
of Schwarzschild metric. This introduces the following fitting parameters: 
r$_{in}$ and r$_{out}$: inner/outer emitting radius of the annulus, 
$\beta$: radial emissivity parameter of the disc and $i$: inclination angle. 
We fitted all the spectra also with a standard cold disc reflection
model (PEXRAV in XSPEC, \cite{aderosa-C2:pex}. 
The best fit parameters listed in Table \ref{aderosa-C2_ironlinetable}) 
in order to:
(a) separately determine the Fe K$\alpha$ parameters and 
(b) check the relative importance to employ the ionized disc model.
All the SAX data were fitted using the XSPEC package 11.1. All the quoted 
uncertainties correspond to 90\% confidence interval for one interesting 
parameter ($\Delta\chi^2$=2.71).
Each model tested was multiplied by a normalization constant
in order to take into account possible miscalibrations between the different 
instruments. We allowed the PDS/MECS normalization to vary between 
0.77 and 0.95, while the LECS/MECS 
normalization ratio was running 
between 0.7 and 1 (\cite{aderosa-C2:cook}).

\begin{table*}
\caption{Results II: The iron line.}
\begin{flushleft}
\begin{tabular}{l c c c c c}
\noalign{\hrule}
\noalign{\medskip}
Source & \bf $\Gamma$ & E$_{Fe}$ & $\sigma$ or r$_{out}$ & EW & $\chi^2$/dof \\
 & & (KeV) & (keV) or (r$_g$) & (eV) & \\
\hline
($^1$)Mkn~335 & 2.04$\pm^{0.17}_{0.03}$ & $^\star$6.4 &6.1$\pm^{3.2}_{0}$ & 660$\pm$150 &126/123 \\
Mkn~841 & 2.12$\pm^{0.10}_{0.02}$ & $^\star$6.4  & 9.0$\pm^{6.3}_{3.0}$ & 260$\pm^{120}_{80}$ & 120.8/119  \\
NGC~3783 & 1.87$\pm^{0.06}_{0.05}$ & 6.39$\pm$0.09 & 0.36$\pm^{0.09}_{0.08}$ &190$\pm$47 & 100.7/102 \\
NGC~4051 & 1.65$\pm$0.10 & 6.2$\pm^{0.2}_{1.5}$ & $<$1 & 135$\pm$200 & 102.3/106  \\
Mkn~509 (1998) & 1.97$\pm$0.03 & 6.88$\pm${0.01} & $^\star$0.4 & 115$\pm$65 & 101.3/141 \\
Mkn~509 (2000) & 1.75$\pm$0.06 & 6.57$\pm^{0.28}_{0.54}$ & 0.55$\pm^{2.0}_{0.18}$ & 149$\pm$100 & 106/126  \\
\hline
\end{tabular}
\end{flushleft}
\small{$^\star$ Parameter fixed at that value in the fit.}
\newline 
\small{$^{(1)}$ The Fe line profile is modelled adding to the ralativistic 
disc line a second narrow gaussian component at E=6.4 keV.}
\label{aderosa-C2_ironlinetable}
\end{table*}
\begin{figure*}[t]
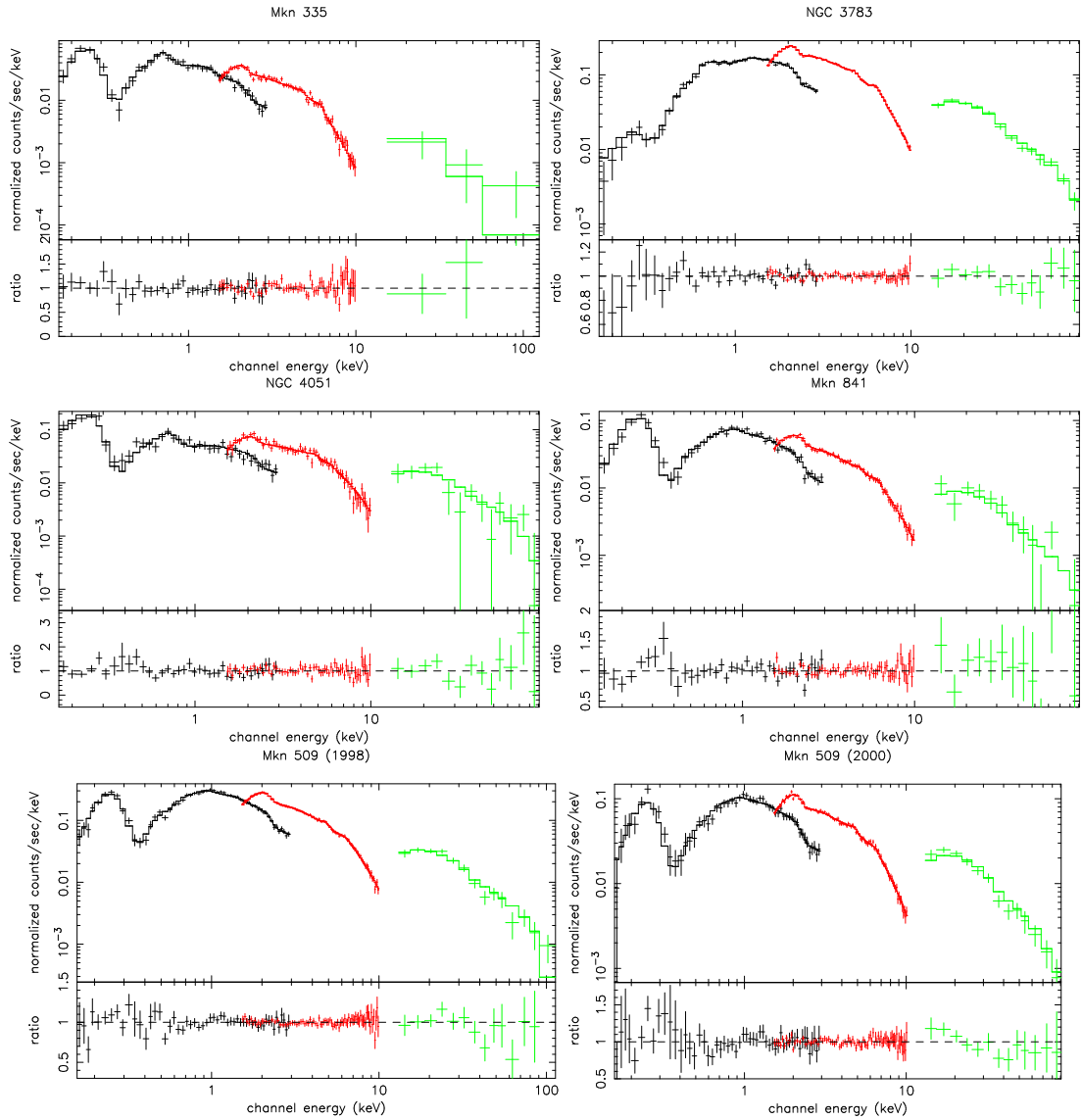
 
\vspace{10pt}
\begin{center}
{\epsfig{file=aderosa-C2_fig1.eps,width=5.0cm,angle=-90}}
{\epsfig{file=aderosa-C2_fig2.eps,width=5.0cm,angle=-90}}
{\epsfig{file=aderosa-C2_fig3.eps,width=5.0cm,angle=-90}}
{\epsfig{file=aderosa-C2_fig4.eps,width=5.0cm,angle=-90}}
{\epsfig{file=aderosa-C2_fig5.eps,width=5.0cm,angle=-90}}
{\epsfig{file=aderosa-C2_fig6.eps,width=5.0cm,angle=-90}}
\end{center}
\caption{In each plot we show LECS (black), MECS (red) and PDS (green) data 
(upper panels) and data/model 
ratio (lower panels) for the ionized disc reflection model 
(the best fit parameters are listed in Table \ref{aderosa-C2_iondisctable}). 
In the case of NGC~3783, Mkn~335 and NGC~4051 the spectral analysis must take 
into account the complex soft X-rays spectra characterized by 
warm absorber(s). 
These features are included in the best fit model 
(see discussion in Section \ref{aderosa-C2_discussion}).
The soft excess observed at E $<$ 2 keV (particularly strong in the 
two NLSy 1s NGC~4051 and Mkn~335), can be accounted by the disc emissivity.}
\label{aderosa-C2_ldratio}
\end{figure*}

\section{Results: summary and conclusions}
\label{aderosa-C2_discussion}
\begin{itemize}

\item

The RF ionized disc reflection model provides an adequate fit for  
all observations examined.
When a cold disc reflection model in employed to fit the spectra 
the general trend is a higher value of the spectral index
and, in addition, a soft X-ray emission component 
(due to the disc emissivity) in required.

\item
In the case of NGC~3783, Mkn~335 and NGC~4051 
the complex low energy spectra (characterized by warm absorber(s) 
and strong soft excess) require a more detailed study. 

The complex structure of the warm gas in NGC~3783 was investigated
in the Chandra observation (\cite{aderosa-C2:kaspi2}).
The model consisted of two absorption-emission components with the same column 
density $\lg(N_H)=22.1$, and differing each other an 
order of magnitude: $\lg(U)= -0.745$, and $\lg(U)= -1.745$.

With the BeppoSAX-LECS energy resolution the warm absorber  
is fitted through the detection of the deep OVII and OVIII K-shell absorption 
edges produced in the low ionization gas.
We added to the ionized disc reflection model in NGC~3783
two absorption edges at energies 0.74 keV ($\tau$=0.71) and 0.87 keV 
($\tau$=0.53) respectively, coincident with the OVII, OVIII K-shell edges 
energy, and a third absorption feature at ~1 keV probably
due to a blend of many iron L-shell absorption lines originated in the second
more ionized gas (\cite{aderosa-C2:kaspi2}).
In the best fit model we included also a narrow Fe line with the intensity 
fixed to the value observed by Chandra 
(I$_{Fe- narrow}$=6.6 $\times$ 10$^{-5}$ photons cm$^{-2}$ s$^{-1}$, 
\cite{aderosa-C2:kaspi2}).
As shown in Table \ref{aderosa-C2_iondisctable} the fit for the ionized 
disc reflection model is good ($\chi^2$/dof=104.2/101).

In the case of Mkn~335 the best fit model in the whole BeppoSAX energy range 
includes a warm absorber component  characterized by OVII and OVIII 
absorption edges ($\tau$=0.67 and $\tau$=0.53 respectively).
The strong soft excess observed in the low energy spectrum 
can be accounted by the emission from a highly ionized disc ($\lg(\xi$)= 2.9, 
see Table \ref{aderosa-C2_iondisctable}). 

NGC~4051 has a warm absorber well known for its spectral complexity
(\cite{aderosa-C2:guainazzi96}). 
Nevertheless the BeppoSAX low energy spectrum was well fitted adding to the
RF model an absorption edge at energy 0.88 keV ($\tau$=0.74). 
Also in this case the strong soft excess can be accounted  by the emission 
from a highly ionized disc ($\lg(\xi$)= 2.65, 
see Table \ref{aderosa-C2_iondisctable}).

We stress here that Mkn~335 and NGC~4051 are classified as a 
Narrow Line Seyfert 1 galaxies, and in this class of sources high 
ionized disc are often observed (\cite{aderosa-C2:ballantyne01a}).

\item

The RF model applied on a sample of NLSy 1s observed by ASCA 
(\cite{aderosa-C2:ballantyne01a}), shows that these 
sources might posses an ionized disc.  
Most of the Sy 1s we analysed observed by BeppoSAX show evidence of 
an ionized disc.
The ionization parameters are anyway much less that that observed in NLSy 1s,
not surprisingly as in latter class of objects are commonly believed 
to accrete at high rates.
NGC~3783 is the ``pure Sy 1'' in our sample that seems to be characterize 
from a highly ionized accretion disc ($\lg(\xi$)= 2.73).

\item
All the sources well reproduced by a RF model are characterized by a 
face-on disc with $i$ $<$ 40$^\circ$ (consistent with the ASCA observations of Sy 1s, 
Nandra et al. 1996) and  with radial emissivity parameter  $\beta$=-3.

\item
The flux variation of a factor of 2 detected between the two observations 
of Mkn~509 (taken 2 years apart), can be completely explained with 
a change of the intrinsic slope $\Delta\Gamma$ $\sim$ 0.2.  
This behaviour is fully consistent with that observed in other 
Sy 1s (NGC 5548, \cite{aderosa-C2:nic00}; 
NGC 4151, \cite{aderosa-C2:piro02}; IC 4329a, \cite{aderosa-C2:done00}; 
NGC 7469, \cite{aderosa-C2:nandra00}; MCG-6-30-15, 
\cite{aderosa-C2:vaughan01}; NGC 3783, 
\cite{aderosa-C2:me}), and with the F$_x$ vs $\alpha$ relationship 
expected in a Comptonization model for the intrinsic continuum emission 
(\cite{aderosa-C2:pop00} and references therein).

\end{itemize}

We are extending this analysis to a wider sample (20 objects) of 
sources observed by BeppoSAX.

\begin{acknowledgements}
The BeppoSAX satellite is a joint Italian-Dutch program.
We thank the BeppoSAX SDC for assistance.
ADR acknowledges G. Matt for useful discussions and financial support from the 
European Association for Research in Astronomy (EARA) Marie Curie
fellowships.

\end{acknowledgements}

\end{document}